\documentclass[a4paper,12pt]{article}
\usepackage{a4,amssymb,bbm,latexsym,mathrsfs,rotating}
\usepackage{bk0,listings,bk1,bm,url,epsfig,xcolor}
\usepackage[mathscr]{euscript}
\usepackage[normalem]{ulem}
\usepackage[english]{babel}
\usepackage{geometry,authblk}   
\usepackage{epic}
\usepackage[reqno]{amsmath}
\usepackage[all]{xy}
\usepackage{graphicx}
\usepackage{color}
\renewcommand\int{\mathbb{Z}}

\renewcommand\ncite[1]{{\cite{#1}}}
\renewcommand\hb[1]{{{#1}}}

\newtheoremit{hypothesis}[definition]{Hypothesis}
\newcommand\bhyp{\begin{hypothesis}}
\newcommand\ehyp{\end{hypothesis}}

\newcommand\eps[2]{\epsfig{file=./#1.eps,scale=#2}}
\newcommand\epss[3]{\epsfig{file=./#1.eps,height=#2cm,width=#3cm}}

\newcommand\angry{\raisebox{-.2em}{\eps{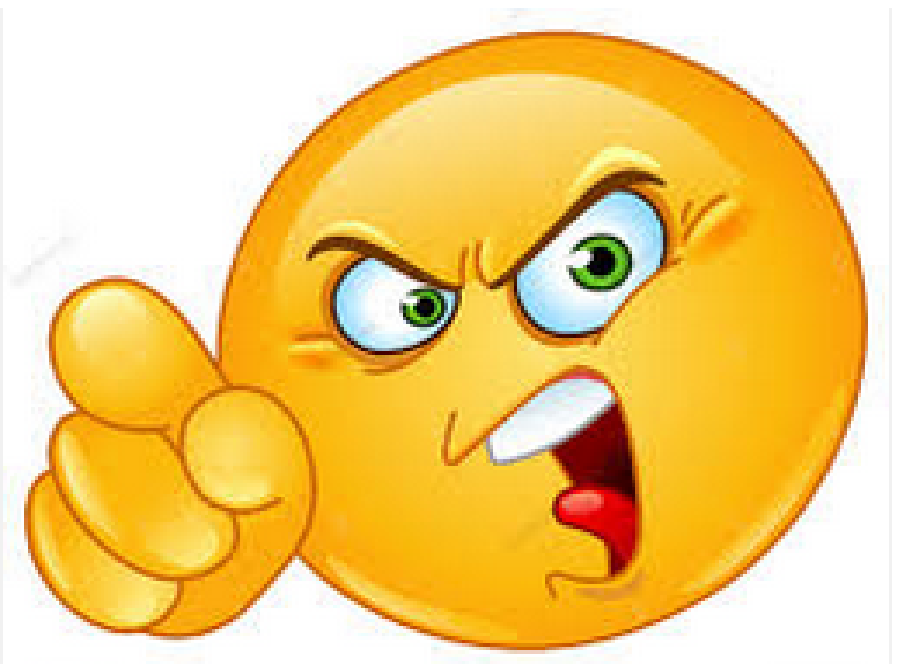}{.07}}}

\newcommand\bax{\begin{axiom}}
\newcommand\eax{\end{axiom}}
\renewcommand\c{{\bf c}}
\renewcommand\T{{\bf T}}
\newcommand\bi[1]{\hb{\emph{#1}}}
\newtheorem{summary}[definition]{Setting and axioms}
\newcommand\bsum{\begin{summary}}
\newcommand\esum{\end{summary}}
\newtheorem{setting}[definition]{Setting}
\newcommand\bset{\begin{setting}}
\newcommand\eset{\end{setting}}
\renewcommand\type{{\mathbb T}}

\renewcommand\S{{\cal S}}
\newcommand\w{{\bf w}}
\newcommand\cC{{\cal C}}

\newcommand\bqn{\begin{equation}}
\newcommand\eqn{\end{equation}}

\let\realtime\time
\newcommand\todays{
  {\mbox{{\small\ {Version $\the\day.\the\month.\the\year:\the\realtime$ (minutes after midnight)}}}}}

\title{Axiomatizing consciousness\\
{\large with applications}}
\author[1]{Henk Barendregt\thanks{Corresponding author.
    Email: henk.barendregt@ru.nl}}
\author[2]{Antonino Raffone\thanks{Email: antonino.raffone@uniroma1.it}}
\affil[1]{{\normalsize Radboud University, Nijmegen, The Netherlands.}}
\affil[2]{{\normalsize Sapienza University, Roma, Italy.}}
\date{\todays}

\usepackage[square]{harvard}
\begin{document}
\bibliographystyle{harvard}

\maketitle

\begin{abstract}
  \noi Consciousness will be introduced axiomatically, inspired by
  Buddhist insight meditation and psychology, logic in
  computer science, and cognitive neuroscience, as consisting of a
  stream of \emph{configurations} that is \emph{compound, discrete},
  and (non-deterministically) \emph{computable}. Within this context
  the notions of self, concentration, mindfulness, and various forms
  of suffering can be defined. As an application of this set up, it
  will be shown how a combined development of concentration and
  mindfulness can attenuate and eventually eradicate some of the forms
  of suffering.
\end{abstract}

\section{Towards consciousness}\label{quest}
Studying phenomena in the `external world' by making conceptual models
has led to physics. Its success gives the impression that also the
human mind could be studied similarly, answering questions like ``How
does consciousness (experience) arise?''  There is, however, a
persistent `explanatory gap' between models of the universe and
`first-person' awareness. This gap is called the `hard problem'
\cite{chal95}. Whatever model of consciousness is proposed, the
question ``And where is awareness in all of this?'' cannot be bypassed
\cite{bitb08}.  Not only is the consciousness problem hard to solve,
it even seems impossible to properly state it\footnote{Personal
communication by Bill Phillips.}.

In contrast to the third person description of consciousness, the
phenomenological approach employs a first person perspective, in which
the experience of consciousness comes prior to anything else. In this
view, matter and the whole universe derive from consciousness as a
construction of the world with predictive value. But then another
problem pops up: ``Why does the external world gives the impression to
be stable?''  \cite{hutshep96}. In this paper the hard consciousness
problem will not be discussed as such. See \cite{weis14,slor15} for
recent discussions. We position ourselves among the phenomenologists:
there is the experience of phenomena that can be studied
phenomenologically. In this way consciousness will be described as an
objective personal phenomenon, not from the brain side, but from the
other side of the explanatory gap: direct experience. The description
will be called objective, since it is claimed that the description is
universally valid, and personal, since it takes place in the mind of a
given person.

The difficulty of defining what consciousness is will be dealt with by
the methodology of the axiomatic method \cite{aris50}. In a given
setting there are \emph{primitive (undefined) objects} (also called
\emph{concepts}, as the objects are mental) and axioms about these
that are taken to be valid. In this way, following \cite{hilb00}, the
axioms form an \emph{implicit definition}\footnote{In planar geometry
  one has as setting that there are points and lines, and that there
  is a relation ``\emph{point $P$ lies on line $l$}'', in notation
  $P|l$. In this setting an example of an axiom is
$$\emph{\mbox{For distinct points $P, Q$ there is exactly one line $l$
      such that both $P|l$ and $Q|l$.}}$$ What actually is a point and
  a line doesn't matter, as long as the axioms are valid for
  these. Since the axioms do not always fully determine the objects,
  one better speaks about an `implicit specification' of the primitive
  concepts.}  of the primitive objects. In the next sections a setting
and axiomatization of consciousness will be proposed using the notions
object (input), state, and action (output) of
consciousness.\footnote{This paper is a continuation of
  \cite{bareraff13}. Another axiomatic approach to consciousness is
  Integrated Information Theory (IIT) \cite{tono12}.  That theory also
  contains the triples object-state-action (using different
  terminology). The model IIT diverges from ours, wanting to propose a
  solution to the hard problem of consciousness.  Although
  \cite{bitb08} argues convincingly that this is impossible, IIT is an
  interesting further analysis of the mechanisms needed for
  consciousness.  Our axiomatization focuses on several applications,
  mentioned in the abstract and detailed below. Further comparison
  between IIT and our model is beyond the scope of this paper.} The
details are inspired by Buddhist psychology, the Abhidhamma
\cite{anur1200}, translated into the language of science: cognitive
neuroscience, mathematical logic and computability. Intended is an
axiomatization of those aspects of consciousness that are shared by
adult humans in possession of their ordinary faculties. The
axiomatization will not touch the hard problem, but aims at describing
certain aspects of consciousness to arrive at some applications in the
domain of computability, learning and deconditioning, and the cause
and eradication of existential suffering.

\section{Consciousness as discrete, quasi-deterministic actor}\label{AC}
\subs{Change}
Science doesn't know what is consciousness. But we know.
Consciousness consists of phenomena, called \emph{configurations} and
are members of a space $\cC$, that change in time.  We write $c_t$ for
the configuration at time $t\in\T$, to be thought of as `what is
perceived at moment $t$'.  Time is not to be seen as a given from the
outside, but as a construct from the phenomena themselves. Time has
passed from $t$ to $t'$ if there is a change from $c_t$ to $c_{t'}$
and there is memory part of $c_t$ within $c_{t'}$. This is called the
\emph{primordial intuition of time}, \cite{brou52}.

The changing configurations create the \emph{stream of consciousness},
which is the function $\c\colon\T\to \cC$ that assigns to a moment $t$
in time the configuration $c_t$:
\bqn\tag{2.1}
\fbox{$\c(t)=c_t$, with $t\in\T$.}\eqn
The stream of consciousness $\c$ may seem like a dynamical system that
changes in time, in which a future state is determined by the state at
present.  Examples of such systems are the following.  1. A single
planet orbiting a star. 2. Conway's Game of Life.

\subs{Actors in a world} But (the stream of) consciousness is not a
dynamical system. The configurations are enacted in an environment,
the world. This way the environment is being changed, which in turn
has an influence on $\c$. Thus consciousness may be better compared to
one planet among other ones in the gravitational field of a star and
the (other) planets. For example the orbit of Uranus could not be
explained by the laws of mechanics w.r.t\ the sun alone: it had an
aberration that only could be explained by the existence of a
hypothetical further planet. In this way the planet Neptune was
discovered. The mathematics involved is becoming complex: the three
body problem (c.q.\ predicting the movements of Uranus and Neptune
with respect to the sun) has chaotic solutions.

\renewcommand\a{{\bf a}}
An \emph{agent} $A$ living in a \emph{world} $W$ consists of the
following.  Both $A$ and $W$ consists of changing configurations;
those of $A$ are denoted by $c,c',c'',c_0,c_1,\ldots$ and similarly
those of $W$ by variations of the letter $w$. Agent $A$ in
configuration $c$ enacts with the world $W$ in configuration $w$. This
enacting is denoted by $c|w$, thereby changing both
configurations\footnote{Dynamical systems are a special case, having a
world that doesn't change (e.g.\ Conway's game of Life).  On the other
hand an agent and its world can be considered as a pair, forming a
single dynamical system. The choice is pragmatic.}. It may be
postulated that the present configuration of agent and world, say
$(c_0,w_0)$ determine both future configurations $(c_t,w_t)$ in a
near-deterministic way\footnote{This postulate is trivial, because any
function is near-computable. A better view is that some aspects of the
future are computable and others not.  These latter aspects depend on
non-deterministic (ND) factors (like the throwing of the dice in the
game of Goose Board). Whether these factors are essentially
non-deterministic or only illustrate a lack of knowledge is irrelevant
(ontological non-determinism vs not knowing non-determinism). In this
situation it is important, for humans and other species alike, to be
able to make educated guesses about the probability of events, as has
been emphasized by \cite{fris10}.  Full knowledge may be desirable,
but it is not feasible.}.  The resulting combined stream of the agent
$A$ thrown in the world $W$ will be denoted as $(\c,\w)$ so that for
$t\in\T$ one has
\bqn\tag{2.2}\fbox{$(\c,\w)(t)=(c_t,w_t)$.}\eqn

\subs{Discreteness of time}
Based on insights of neurocognition, micro biology and vipassana
meditation we postulate that time is \emph{discrete}. This means that
$\T$ is not modeled by the set $\real$ of real numbers, but by
$\int=\set{\ldots,-2, -1,0,1,2,\ldots}$ the set of integers. So
\bqn\tag{2.3}\fbox{$\T=\int$.}\eqn
This explains how for agent \bi{$A$} in world \bi{$W$} the combined
stream $(\c,\w)$ develops by the repeated interaction operation
$c|w=(c',w')$, as a ND-computable function:
\[\fbox{\xymatrix{ &&c'\ar@{.>}[rd]&&c''\ar@{.>}[rd]\\
\ldots&c|w\ar@{->}[ru]\ar@{->}[rd]&&c'|w'\ar@{->}[ru]\ar@{->}[rd]
&&c''|w''&\ldots\;,\\ &&w'\ar@{.>}[ru]&&w''\ar@{.>}[ru] }}\eqno{\raisebox{-3.4em}{(2.4)}}\]
creating \bi{streams} $\hb{\c\colon c\to c'\to c''\to\ldots}$ and
$\hb{\w\colon w\to w'\to w''\to\ldots}$ of configurations and states
of the world. The $\w$ could be called the \bi{trace} or
\emph{footprint} of the agent in the world.  The transitions from the
interacting $c$ and $w$ to $c'$ and $w'$ take place in \bi{discrete
  time}, that imaginatively could be called \emph{stroboscopic}.  This
creates phenomenological time.  We have chosen $\T=\int$ and not
$\T=\nat=\set{0,1,2,\ldots}$ to make time without beginning. The
reader may like to make another choice.

In \cite{zylbetal11} it is explained that discreteness of the stream
of consciousness neatly answers the question of von Neumann how it is
possible that the human mind, being based on a biological substrate
with its inherent imprecision, is capable to arrive at the precision
that is available in e.g.\ mathematics. This is similar to a digital
CD that represents sound with less noise than an analogue record.

\subs{Stream of consciousness is ND-computable} The stream of
consciousness proceeds in mutual dependency on the stream of the
world. The progression is determined by repeatedly applying the the
operation $c|w$. In this way one obtains a new pair of configurations
$(c',w')$ that are being subject to their interaction
$c'|w'$. Etcetera. We assign the task of obtaining the next $c'$ or
$w'$ to the agent $A$ and its world $W$; so we have
\bqn\tag{2.5}
\begin{array}{|rcl|}
\hline
A(c,w)&=&c'\hoogm;\\
W(c,w)&=&w'.\\[.1em]
\hline
\end{array}
\eqn
That is $c|w=(A(c,w),W(c,w))$. The functions $A,W$ with \bqn\tag{2.6}
\fbox{$A\colon \A\times \W\to\A,\;W\colon\A\times \W\to\W$}\eqn are
postulated to be ND-computable, i.e.\ computable by non-deterministic
Turing Machine. The non-determinism is caused by the following.
1. There are neural nets in the brain of a human agent that act
adequately but not with 100\% precision; 2. not knowing how the world
reacts; 3. not knowing what other agents are doing to the world. This
third point can be seen as part of the second.

As motivation for the axiom of ND-computability of the stream of
consciousness one can refer to: functioning of neurons, see
\cite{maasmark04}. The Buddhist view, and corresponding meditation experience,
that everything has a cause (dependent origination) also motivates
this axiom. The axiom also is consistent with the Turing Thesis
\cite{turi37} that states that human computability is exactly machine
computability.

Summarizing. Consciousness is a quasi-deterministic actor, where the
non-determinism is caused by the imprecision of the agent and the
unknown aspects of the world. Nevertheless, because the actions are
digitized, great precision is possible.

\section{Compound consciousness}
\subsubs{Input, state, action}
Acting in a world is made
efficient by sensors, channels for input ($i$), and actuators, for
action ($a$). Behaviorism took as position that humans could be
described by the set of pairs $(i,a)$ (in short $ia$), also called
`stimulus and reaction'. In this line of thinking one could write
\bqn\tag{3.1}
\fbox{$\c(t)={c_t}={i_ta_t}$, with $t\in\int$}\eqn
This, however, is a limited view, as a person doesn't behave in the
same way if being subject to the same input. Therefore next to $i$ and
$a$ one needs an (internal) `state' $s$ to describe the agent.  This
`mind-state' $s$ can be considered as `the tendency to act in a
certain way'. This results in postulating that for the configurations
$c$ of an agent $A$ one has $c=isa$, so that the stream of
consciousness $\c$ can be considered to consist of three
streams\footnote{This is how the transitions in a Turing Machine can
be seen. The Read/Write device (R/W-head) is positioned on a cell and
reads $i$. Then depending on this and on the state $s$ an action is
performed: either moving the R/W-head, or writing a symbol on the
cell where the R/W-head is positioned, or changing the inner state.}.
\bqn\tag{3.2}{\small 
{\fbox{$\hoogm
    \c\;=\;\ldots\to\underbrace{i_{-1}s_{-1}a_{-1}}_{c_{-1}}\to\underbrace{i_{0}s_{0}a_{0}}_{c_{0}}\to
    \underbrace{i_{1}s_{1}a_{1}}_{c_{1}}\to\ldots\;=\left\{\begin{array}{l}\ldots,
i_{-1},i_0,i_1,\ldots\\ \ldots, s_{-1},s_0,s_1.\ldots\\ \ldots,
a_{-1},a_0,a_1,\ldots \end{array}\right.$
}}}\eqn

\subsubs{Feeling tone: reward system}
For humans (and other species)
it is useful to make a further division. 1. Writing $s=s^fs^c$, where
$s^f$ is the feeling tone and $s^c$ is the rest of the state of
consciousness. The $s^f$ is an element of $\set{--,-,0,+,++}$ and
indicates whether the present configuration is felt as very
unpleasant, unpleasant, neutral, pleasant, very pleasant. It is the
reward-punishment for humans and other species; nature makes certain
things pleasant, like eating and making children, in order to make
Homo Sapiens thrive.

{\subsubs{Cognition: memory, language, mental programs}
Another subdivision, notably for humans, is to add a group $i^m$ for
`cognition'\footnote{Traditionally this is called the group of
  `perception'.}, consisting of concepts and images and split $i$ as
follows: $i=i^bi^m$. The objects of $i^b$ consist of input from the
physical senses, hence the superscript `$b$' refering to `body'. The
objects of $i^m$ consist of mental images, concepts, and intentions to
act.  Except for pathological cases, humans can distinguish these
respectively from actual input through $i^b$ and from actual execution
of the intended act as $a$.

  The elements of the streams in $c=i^bi^ms^fs^ca$ are acting in an
  associative way. The sound of a bell ($i^b$) preceding a meal for a
  dog that triggers saliva, after a couple of times is enough to
  trigger the saliva without a meal. In general associations between
  elements of the $isa$ may trigger occurrences of other objects
  possibly in another stream. The group $i^m$ has a rich potential of
  elements that can be triggered by an event coming in through $i^b$,
  and causing in its turn the right reaction in $a$.

  For this to work well there is \emph{cued recall}. After a
  particular object $o_1$ in say $i^b$ is presented several times and
  followed by another object $o_2$, the presentation of just $o_1$ may
  trigger the memory of $o_2$. In a small brain cued recall has
  limited reliability (the recalled $o_2$ may not be correct) and
  capacity (only a limited numbers of pairs $(o_1,o_2)$ may be
  stored. This limitation can be increased considerably,
  \cite{deBr03}, at the cost of brain tissue and energy
  consumption. In this way Language and mental programs can be
  developed. }

\subsubs{The five groups} Taken together one obtains the five groups,
aka aggregates/skandhas:
\bqn\tag{3.3}\fbox{$c=\underbrace{i^bi^m}_{i}\underbrace{s^fs^c}_{s}a\hoogm$,}\eqn
so that the stream of consciousness has five substreams. The new
substreams \bqn\tag{3.4}\fbox{$\begin{array}{l} s^f=\ldots \to s^f_{-1}\to
  s^f_{0}\to s^f_{1}\to\ldots\\ i^m=\ldots \to i^m_{-1}\to i^m_{0}\to
  i^m_{1}\to\ldots \end{array}$}\eqn are the stream of feeling tones and that
of mental activities, like thinking or imagining. These two streams
often are being hypertrophied (in the sense of getting much attention)
in human existence, notably reinforcing each other.

\subsubs{Finer details of consciousness} A triple $c_t=i_ts_ta_t$ (or
more accurately a quintuple $c_t=i_t^bi_t^ms_t^fs_t^ca_t$) is called a
\emph{ceta} (aka \emph{citta} or \emph{mind-moment}). A state can be
approximately seen as a large array of values (parameters). Think of a
possible state of the weather, e.g.\ a local snowstorm.  Relevant for
that state are the temperature, humidity, wind, and more at the
different relevant local positions. In Buddhist psychology, the
Abhidhamma, the mind-state $s$ is seen as such an array of many so
called mental factors, called \emph{cetasikas}.  As feeling tone $s^f$
is such an important factor, that is always present, it is singled out
in the five groups. Other mental factors, that however are not always
present, are \emph{aversion, desire} on the unwholesome side, and
\emph{mindfulness}, to be introduced below, and \emph{compassion} on the
wholesome side.

\hb{\section{Self} That an agent in the world proceeds with a
  ND-computable stream of consciousness may be expressed by saying
  that it is `impersonal'. It just follows the laws of nature,
  depending on the configuration of $A$ and the state of the
  world. Another way of expressing this is by saying that $A$ is
  self-less. It proceeds without independent existence, just like like
  a glider crawls diagonally over the field of Conway's Game of Life,
  or like a wave towards the shore, that seems to proceed from a
  pebble thrown into the middle of a pool. In the latter case water
  only moves up and down, not sideways, as becomes clear when placing
  a ping-pong ball in the water. Nevertheless within the life-stream
  of the agent it can happen that a self is being formed. It is a
  dynamical process consisting of a collection of behavioral
  strategies that protect and take care of the individual. This self
  needs some balance: fine tuning of the different sub-strategies.
  
  \subs{Healthy attachments} When homo sapiens considered as agent
  grows up it learns as a baby first the following: relating $a$ and
  $i$, so that some control over the environment can be
  obtained. Shortly after in the development of a child, as each $i$
  is coupled with $s^f$, the actions will be directed towards avoiding
  input with unpleasant $s^f$.

With the capacities so
  far: acting towards pleasant input in an intelligent way, learned
  from the social environment, agent $A$ develops strategies
  that are good for $A$, for itself. If this happens in the right way,
  one has developed a healthy self through healthy attachments.
  
  \subs{Selfing} If one doesn't have enough empathy, the capacity to
  imagine the state of others in a given situation, the notion of self
  may become too central and becomes counter productive. If one
  mentions to often `I, me, mine', and acts accordingly, then one will
  be avoided by people in one's environment.

  \subs{Wrong view} The self that has been described as a dynamical
  process is used so often, that it gets reified as a thing. In the
  same way as the wave is seen as an object that moves towards the
  shore, the self is perceived as an entity with independent
  existence. This is called `Wrong View'. In the first place this
  causes fear of death. But many more problems will result, as Wrong
  View creates the idea that one needs to defend self. Also it leads to the
  unwholesome habit of selfing.

}

\section{Mindfulness: mechanism and application (ER)}
\subs{Mechanism of mindfulness} In the given model of consciousness
one can define mindfulness. In this way one primitive term can be
eliminated.\\

\noi \emph{Mindfulness} at $c_{t+1}$ is a mental factor that has (part of) the
previous ceta $c_t$ as object. If $c_t=isa$, then the next ceta being
mindful means that it is $c_{t+1}=(`isa\mbox{'})s'a'$.
One speaks of the `right' mindfulness if $s'$ contains friendliness.\\

Mindfulness can help emotional regulation. Suppose $c_t=i(\angry+s)a$
is a ceta in which the mind-state contains the cetasika (mental
factor) of angriness. The presence of this unwholesome factor makes it
probable that the action $a$ is unwholesome, increasing the chance of
suffering at some or more future consciousness moments.  Being mindful
of the angriness at the next ceta can be seen as
$c_{t+1}=(i+`\angry$'$)sa'$.  The transition
\begin{equation}\tag{5.1}\label{eq:angry}
    i(\angry+s)a\longmapsto (i+`\angry\mbox{'})s'a'
    \end{equation}
is said to be the transformation of \emph{being angry}, possibly with
unwholesome act $a$, to \emph{seeing angriness}, with an
equanimous mental state $s'$ and wholesome act $a'$.

Application of mindfulness: purification. \emph{Mindfulness training}
consists of exercising the transition (\ref{eq:angry}) so that
mindfulness becomes easy to apply.  To increase the effect of
mindfulness in the direction of emotional regulation (ER) one may
train it so that it becomes strong and sharp. Strong means that it is
being applied during a longer time period; sharp means that it is
being applied with a high frequency. In Section \ref{release} we will
see that there is another application of mindfulness, as tool to
insight and release.

\hb{\subs{Mindfulness as risk factor} A strong and sharp form of
  mindfulness is useful for removing counterproductive mind-states.
  When mindfulness has been sufficiently developed, so that it
  possesses a high resolution and can be maintained for an extended
  period, eventually it will show that consciousness is
\begin{equation}\tag{5.2}
  \fbox{compound, fluctuating, impersonal,}
  \end{equation}
and therefore a cause of suffering. In the Buddhist tradition, \cite{budd400},
one mentions the \emph{three fundamental characteristics of existence} (and thereby of consciousness):
\begin{equation}\tag{5.3}
  \fbox{non permanence, suffering, non self.}
\end{equation}
Experiencing this causes further `insights': feelings of (irrational)
fear, delusions of seeing (non existing) danger and (utter)
disgust/nausea, often experienced in quick succession. These form an
impressive cross-section of psychiatric conditions.

}

\section{Suffering}
One can distinguish three essentially different forms of suffering and distress.
\[\btab{lll}
\wit{2}\eps{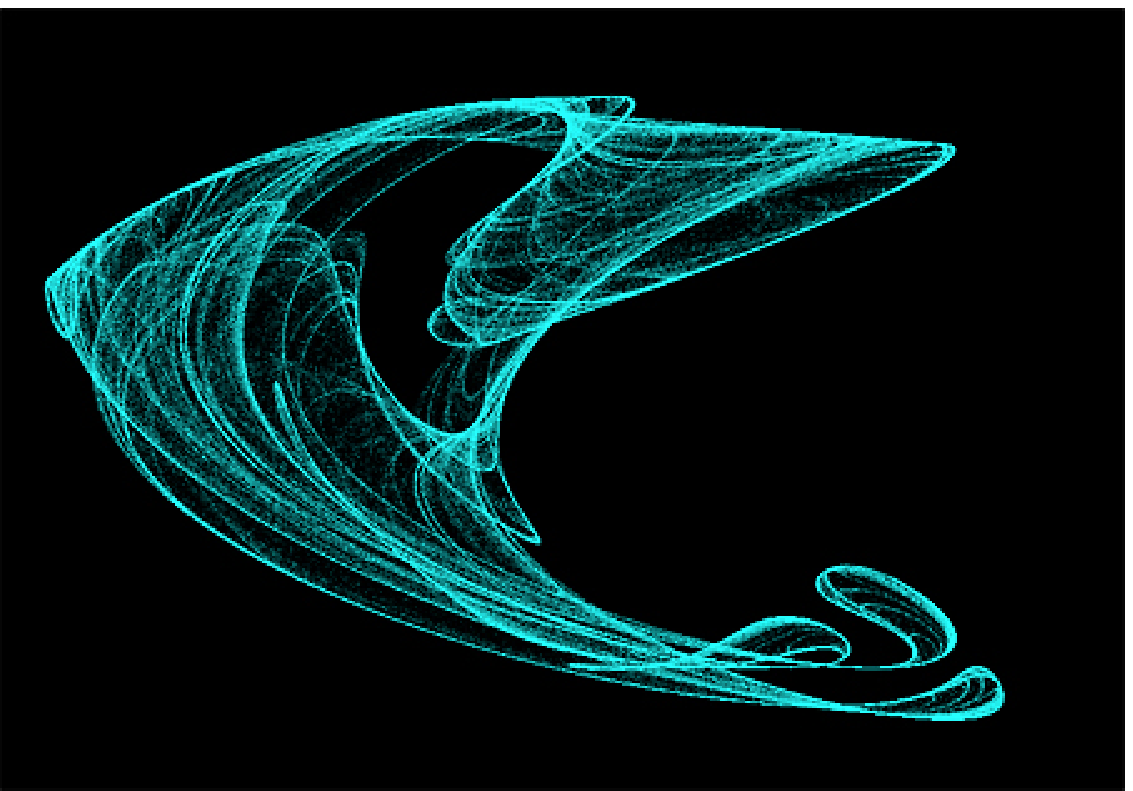}{.30}&
\wit{7}\eps{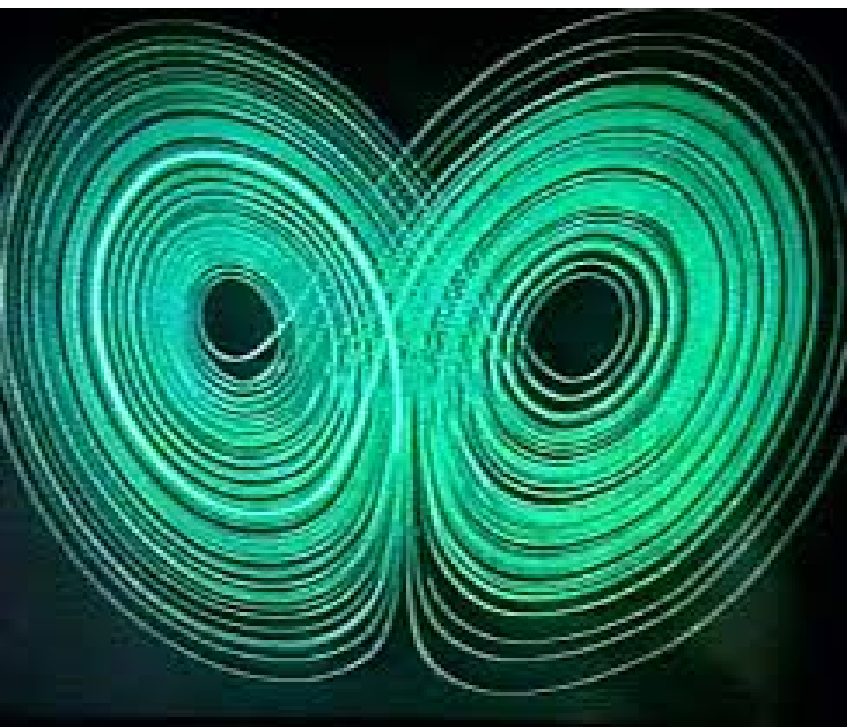}{.33}&
\wit{7}\eps{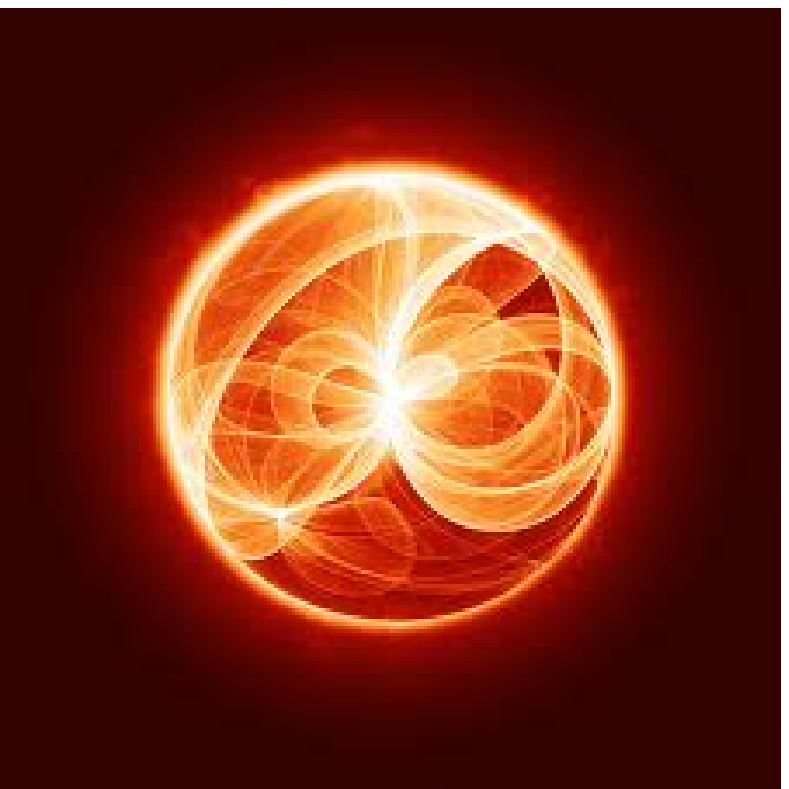}{.30}\\
{{\small \btab[t]{l}1. Distress by\\
avoiding pain
\etab}}&
{{\small \wit{5}\btab[t]{l}2. Distress by\\
avoiding change\etab}}&
{{\small \wit{5}\btab[t]{l}3. Chaos \& Lack:\\
    existential fear
\etab}}
\etab\]

\weg{\[\btab{lll}
\eps{chaotic-mind}{.25}&
\hspace{.29em}\wit{5}\eps{StrangeAttractor}{.27}&
\hspace{.29em}\wit{5}\eps{attractor2}{.25}\\
{{\small \btab[t]{l}Ordinary distress:\\
restlessness\\
avoiding pain
\etab}}&{{\small \wit{5}\btab[t]{l}Cover-up:\\ stable but rigid\\
avoiding change\etab}}&
{{\small \wit{5}\btab[t]{l}Chaos \& Lack:\\
existential fear\\
`no escape'\etab}}
\etab\]}
\subs{Suffering as pain} The most basic form of suffering comes in the
form of feeling-tone $s^f$ having a negative value. Things are unpleasant or even very much so.

\subs{Suffering as change} The strategies constituting the self have
as goal to minimize pain and maximize pleasure.  If one has some
success in this, then one likes to keep the life style one lives. For
that reason change is felt as a threat and is felt as cause of
suffering.

Next to this there is also a mechanism of trying to hold onto one's
lifestyle, even if it is not conductive to decreasing suffering.
This will be explained in the next subsection.

Both the drive to accomplish what one wants and to cover up what one
fears lead to rigidity.

\subs{Suffering from Lack} The fact that consciousness is progressing
as a stream that is compound, fluctuating as a stroboscope, and
impersonal, is a serious blow to self, when there is the Wrong View of
it being permanent and substantial (having independent
existence). Therefore all kind of defense mechanisms create a
cover-up, so that this fundamental fact will not be seen. This
cover-up becomes rigid, if one gets the feeling that it is taken
away. This explains the second reason why change may be felt as
suffering, mentioned in the previous subsection.

If, on the other hand, one doesn't succeed in maintaining the
cover-up, then outright existential fear appears. This fear is not
related to objects, like a wild animal, that appear in the world.  It
is related to the mechanism of consciousness and therefore is
difficult to understand by friends that would like to provide help,
but are unfamiliar to the experience of the three fundamental
characteristics.

\subs{(Un)wholesome actions} An action is (un)wholesome if the chance
of later resulting suffering (increases) decreases.  A mind-state is
(un)wholesome if it leads to (un)wholesome acts.  While hedonist acts
are intended to lead to immediate pleasure, wholesome acts are
intended to lead to sustainably avoiding suffering.

\section{Release: $\down$suffering \& $\up$freedom}\label{release}
To increase resilience against stress and make it sustainable one
needs to release existential suffering. For this the insight
meditation tradition \cite{} has created the triple training:
\bqn\tag{7.1}
\fbox{behavior $\longmapsto$ concentration $\longmapsto$ wisdom.}
\eqn
The development of behavior, also called \emph{discipline} or
\emph{ethics}, is towards having respect for oneself, others, and the
world. This prevents necessary actions in the future and simplifies
life. For example if one doesn't steal one will not risk to come into
contact with the police to be charged for theft. This helps enabling
to develop a lifestyle apt to build concentration, i.e.\ being able to
restrict attention to fewer objects. Details how to do this are beyond
the scope of this paper, but can be found in many meditation manuals,
e.g.\ \cite{maha16}. Then, finally, it becomes possible to obtain insight
into the functioning of our body-mind system so that unwholesome
mental loops (vicious circles) can be defused and avoided.

An important aspect of the training of behavior and concentration is
that also mental activity $i^m$, which is both an action and an input,
decreases.

It is not the case that one first fully develops ethical behavior,
then concentration, and only then insight arises. With some discipline
in behavior, some concentration may be developed, and then some wisdom
arises. With that wisdom one is motivated to increase discipline, so
that concentration and wisdom can be developed further. This then
leads to an upward spiral.

Discipline means that one follows a mental program a plan.
Concentration means that one is able to keep one's attention to a
desired object, the meditation object, for example the physical
sensations of the movements related to breathing. This is practiced by
taking a meditation object with as aim to keep it as long as possible
in focus. Each time when attention has drifted somewhere else, often
without even noticing this, as soon as one is aware of this, one
gently brings attention back to the chosen object. When this is done
continuously, eventually concentration grows and the period to remain
focused on the meditation object increases considerably.

With enough discipline and concentration one is able to restrict the
$i$ and $a$ that they are approximately constant and become $i_0$ and
$a_0$. Then a usual stream of consciousness like
\begin{equation}\tag{7.2}\label{eq:6.2}
\fbox{$\ldots\to isa\to i'a's'\to i''a''s''\to\ldots$}
\end{equation}
becomes
\begin{equation}\tag{7.3}\label{eq:6.3}
\fbox{$\ldots\to i_0sa_0\to i_0s'a_0\to i_0s''a_0\to\ldots$}
\end{equation}
with the input and action fixed to $i_0$, $a_0$, respectively.
This means that the only change is happening in the stream of mind-states
\begin{equation}\tag{7.4}\label{eq:6.4}
\fbox{$\ldots\to s\to s'\to s''\to\ldots$}
\end{equation}
Being for some longer time in this scenario is restful. But certain
tendencies remain present.  After stopping meditation, going back to
sensory and mental input one returns to the usual scenario
(\ref{eq:6.2}).  Nevertheless having felt the quietness of
(\ref{eq:6.3}) is already refreshing, wholesome, and increasing one's
resilience.

But it is possible to develop something better: sustainable resilience.
Not counting mental or sensory input, it can be assumed that
there are only a limited number of mind-states. Therefore the stream
of mind-states will enter a loop:
\begin{equation}\tag{7.5}\label{eq:6.5}
\fbox{$s\to s'\to s''\to\ldots s^{(k)}\to s$}
\end{equation}
If one is fully aware of this loop, or at least of a subloop jumping
now and then a few positions, then habituation occurs and
consciousness occurs without an object arises where even $i_0$
disappears. This is called \emph{nibbana/nirvana}. It causes a
powerful reset, enabling the stream of consciousness to escape from
the quasi-attractor in which it was caught for a long time. Wrong View
becomes Right View, that was already intuitively clear during the
insight of Lack, but it was not yet accepted.

The transitions $(7.2)\longmapsto(7.3)\longmapsto(7.5)$
can be intuitively depicted as follows:
\[\btab{ccl}\eps{StrangeAttractor}{.40}&\epss{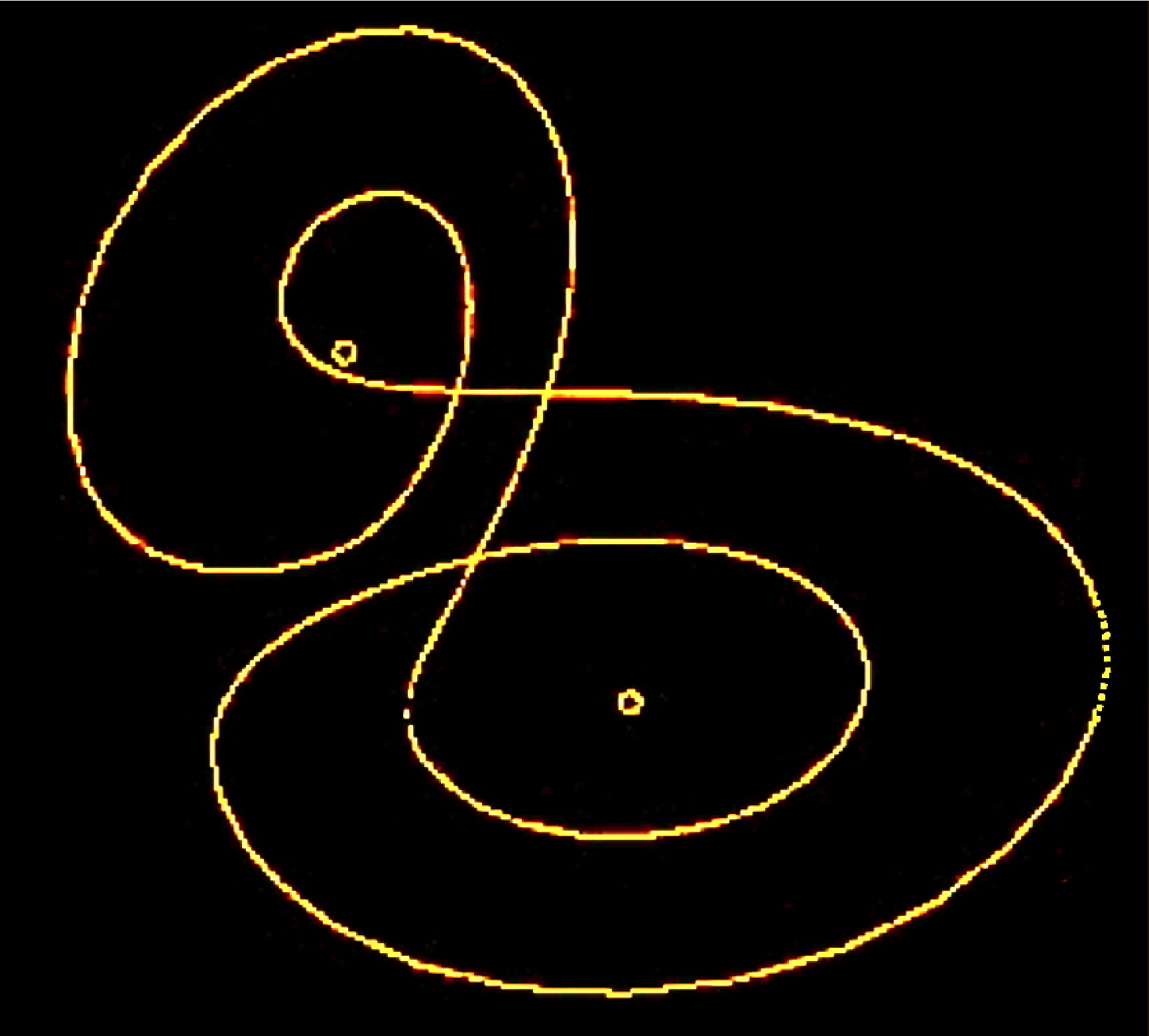}{2.93}{2.9}&
\eps{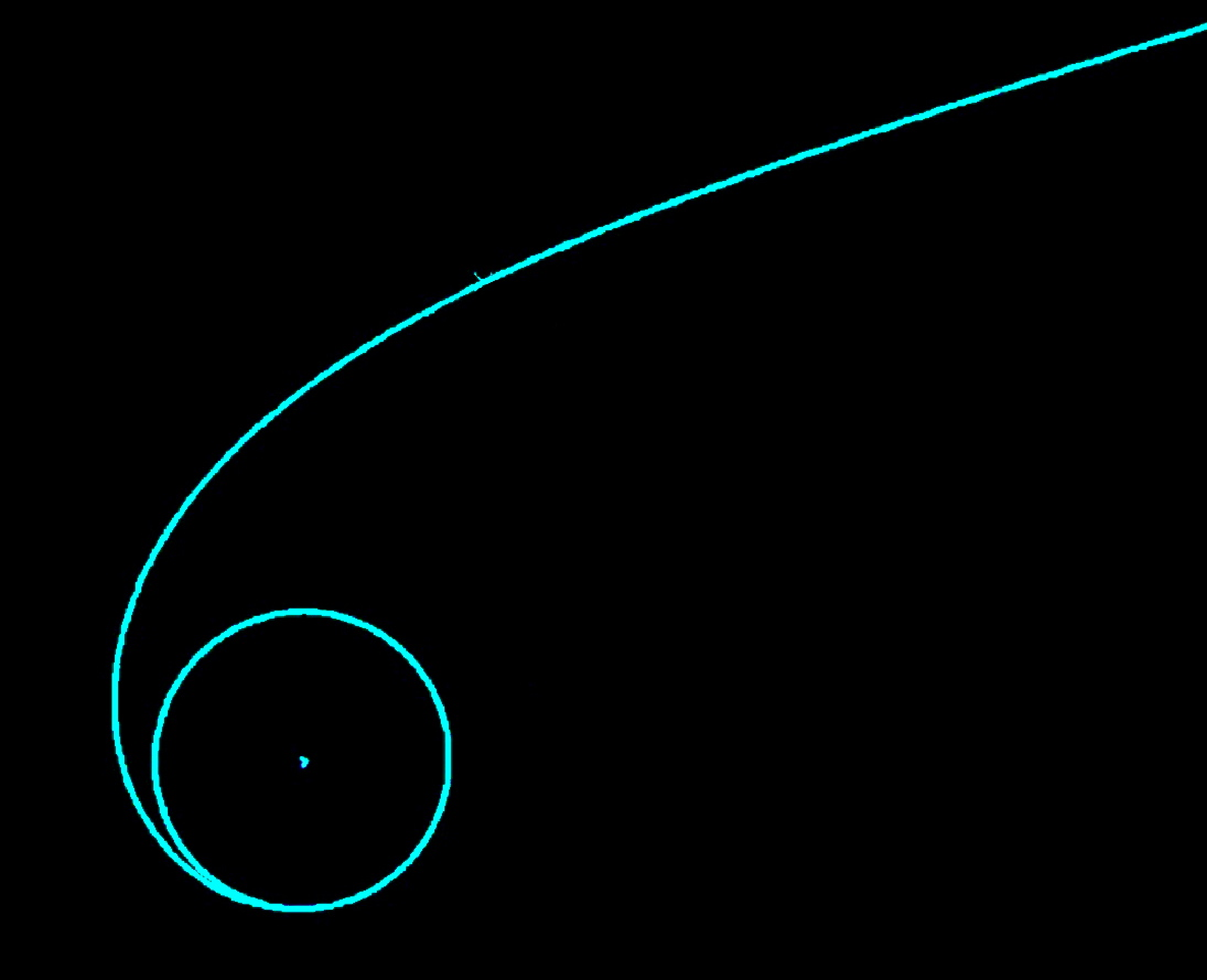}{.093}\hspace{-.2em}{\raisebox{6.575em}{\hb{\hb{$\to$ {\scriptsize \ul{freedom}!}}}}}\\
ego (cover-up)&concentration&\wit{12}release\\[-.2em]
\etab\]

\section{Freedom paradox} There is a remarkable pseudo paradox.  Being
fully aware of the loop (\ref{eq:6.5}) one intuitively understands
what is called `Dependent origination'.  Basically this states that
the stream of consciousness (\ref{eq:6.5}), but then also
(\ref{eq:6.2}), is subject to a quasi-deterministic process. This is
liberating, as one is no longer obliged to pretend one has an
essential say in the propagation of our stream of consciousness. No
longer pretending frees us from rigidity fixated on the self-image we
held on to for a long time.  Therefore there is the freedom paradox:
\bqn\tag{8.1} \fbox{We become free by realizing that we are fully
  determined.}\eqn To understand this, we may compare homo sapiens to
a goat that is attached by a rope around its neck to a pole in the
grass. Consequently the animal can graze only in a circle around the
pole. The goat learns from someone, or invents it auto-didactically,
that to become free one should gnaw on the rope.  When the goat has
succeeded to break the rope, it is free to walk away from the farm
where it is being held, walk into the fields, forests, and mountains
to find other goats for playing and mating. Thereby the goat follows
its way of being conditioned. It even can go back to the farm.  In
this simile the rope for homo sapiens consists of the image one has of
oneself, including our desires and fears. One is attached to this
self-image, in order not to feel the fundamental Lack \cite{loy96} of
self, of substantial independent being. Freedom consists of having
`algorithms' that are pretty good in calculating in an intelligent and
compassionate way what is our best surviving strategy. This way our
actions are based on a flow and no longer on ideas that create our
narrative being. Another way of stating\footnote{Formulation by Karin
  Videc.} the freedom paradox is the following.
\bqn\tag{8.2}\fbox{There is freedom. But it is not ours.}\eqn

Something similar has been stated in \cite{merl13}, in a literary way.\dwn
\emph{I am a psychological and historical structure. Along with
existence, I received a way of existing, or a style. All of my actions
and thoughts are related to this structure, and even a philosopher's
thought is merely a way of making explicit his hold upon the world,
which is all he is. And yet, {I am free, not in spite of} or beneath
{these motivations,}  {but rather by their means}. For that meaningful
life, that particular signification of nature and history that I am,
does not restrict my access to the world; it is rather my means of
communication with it.}\\
\wit{68} {\small Merleau-Ponty$\colon$ \emph{Phenomenology of perception}}

\section{Layers of consciousness} Using our physical
senses and possibly also the mental sense through which the $i^m$
arrive, is overwhelming. Therefore the human mind has a mechanism of
attention that makes a selection.  This can be modeled by allowing
each $i$ to be a large set of values, together with a (chosen) subset
$F\sbs i$ of values to which attention is being paid. In the same way
action $a$ can be seen as a large set of possible actions to which one
needs to apply attention as subset $G\sbs a$, to select the intended
actions.

As we live in a complex environment we are not aware of all the input
stimuli that reach our eyes. We make a choice using attention. So
input $i^b$ in fact is $i^b=\lr{\vec i;F}$, where $F$ is a subset of
the large set of `pixels' $\set{\vec i}$ falling on our visual field,
chosen by attention.

\subs{Forms of consciousness} One can ride over a
well-known bridge in town without realizing that one does
this. Arrived in the other part of town suddenly one realizes `We are
here, so I must have crossed the bridge.' Consciousness is sometimes
described as proto-consciousness plus knowing. As the example shows,
this knowing part is not always there.  In the theory presented so
far this can be modeled as having a (series of) mind-moment(s)
including the mental factor of mindfulness that enables input not via
the physical senses, but more directly from the information of the
previous mind-moment.

One may even differentiate further. Pre-consciousness of an
object $i_0$ may be described as a $((\set{\vec i};F),s,a)$ in which
$i_0$ is among the $\vec i$, but is not attended to, i.e.\ not in $F$.
Proto consciousness of an object $i_0$ is such that $F$ focuses on at
least $i_0$. And as stated, full consciousness arises when $i_0$ is
also observed in the next mind-moment by mindfulness.
\bqn\tag{9.1}\framebox{
  \btab{rcl}
  (full) consciousness& =& proto-consciousness + knowing\\
  proto-consciousness&=&pre-consciousness + attention
\etab}
\eqn
See \cite{hobs09} and \cite{dehaetal06} where these distinctions have
been made, using slightly different terminology.

{\subs{Layers of agents} Conscious agents $A,B$ can be combined by
  diverting the actions of $A$ towards the input of $B$ and vice versa
  the actions of $B$ towards the input of $A$. This has been done in
  an attractive way by \cite{hoff14} and \cite{hoffetal18}.  By also
  considering the physical base as agent interaction, as is done in
  quantum physics, these authors and also \cite{rove21} coin the
  interesting possibility that the explanatory gap of the body-mind
  problem may be bridged.}

\section{Conclusion}
Consciousness is \bqn\tag{10.1}\framebox{compound, fluctuating,
  impersonal.}\eqn Discovering this has strong psychological
implications. It may explain on the one hand part of the psychiatric
phenomena: fear (panic attacks and phobias\footnote{It also has been
  described in \cite[Ch. XIII]{bare82} that phobias appear after one
  has had experience of non-permanence (called `chaos') and non-self
  (called `it'). In this Chapter phobias are described as
  repersonalization after the depersonalization. In \cite{bare96} this
  idea is generalized as the so-called `cover-up' model.}), delusion
(paranoia), disenchantment (depression).  On the other hand that it is
possible to develop the mind in impressive ways.  Through combined
phenomenological and neurophysiological investigations this may
eventually give full insight into the objective nature of
consciousness, its ailments and possibilities.

\section*{Acknowledgments} 
\hb{The Lorentz Institute at Leiden University and the Netherlands
  Institute for Advanced Studies provided support in the form of the
  Distinguished Lorentz Fellowship. The following persons provided
  much appreciated help: Mark van Atten, Martin Davis, Fabio
  Giommi. Wolfgang Maa\ss, Bill Phillips, and Karin Videc.}

\nocite{veenbare10,veenbare15}

\bibliography{br}

\edoc